\documentclass[12pt]{article}
\usepackage{amssymb,amsmath,graphicx,wrapfig}
\usepackage[mathscr]{eucal}

\author{}
\date{}
\begin{document}

\def\p{\partial}
\def\e{\rm e}
\def\i{\rm i}
\def\d{\rm d}
\def\what{\widehat}
\def\Tr{\rm Tr}
\def\rot{\rm rot}
\def\div{\rm div}
\def\grad{\rm grad}
\date{}

\title{Tomographic representation of ECG-signal}
\author{Yu.M. Belousov$^1$, N.N. Elkin$^2$, V.I. Man'ko$^3$, E.G. Popov$^4$ and  S.V. Revenko$^4$ }

\maketitle

\noindent{$^1$\small{ Moscow Institute of Physics and Technology, Dolgoprudny, Moscow reg.}\\
$^2$ \small{State Research Center of Russian Federation 'Troitsk Institute for Innovation and Fusion Research', Moscow, Troitsk, Russian Federation} \\
$^3$\small{ Lebedev Institute RAS, Moscow} \\
$^4$ \small{National Medical Research Cardiology Center, Ministry of Health of the Russian Federation, Moscow}}

\smallskip
Corresponding author: Yu.M. Belousov, e-mail: theorphys@phystech.edu

\bigskip


\section*{Abstract}

Tomographic representation of an arbitrary nature signal is considered. Comparison of a tomographic representation and a fractional Fourier transform is carried out. Application of tomogram representation or identical to it the Radon transform for processing medical ECG signals is illustrated. The first results show that in contrast to routine Fourier analysis, the Radon transform revealed an additional information for  patients with the early signs of ischemic heart disease (IHD) with respect to healthy persons.

\section{Introduction}

In formalism of quantum mechanics the system pure states are described by
the complex  wave function depending on position of a
particle~\cite{shroedin26}. In classical physics the evolution of the
particle state or of any system state is described by the function of time
or some signal, e.g. electric signal associated with activity of human
heart and reflecting properties of its state. Analysis of quantum system
states i.e. the study of wave function  properties  in quantum physics or
analysis of signal functions of time in classical processes  use the
similar mathematical methods. Fourier analysis~\cite{Four}, wavelet
analysis~\cite{wavelet} are examples of such general methods.In quantum
mechanics the different kinds of transforms of wave function or density
matrices~\cite{Landau,von Neumann} describing the states of systems with
thermal fluctuations are used to clarify better the relations  of
classical  and quantum properties of the systems. For example Wigner
introduced~\cite{Wig32} the Wigner function of position and momentum of
quantum particle analogous to the probability density of classical
particle in its phase space. Some other different versions  of such phase
space functions called quasi-distributions were introduced in quantum
mechanics,e.g.,  Husimi-Kano function~\cite{Husimi40,Kano65} and
Glauber-Sudarshan function~\cite{Gla63,Sud63} which are widely used
to describe the properties of laser electromagnetic fields. In view of
Fourier transform of the functions of time the classical signal can be
associated with the function of frequency. In quantum mechanics  the
Fourier transform of the wave function depending on particle position
determines the wave function in momentum representation and this wave
function  depends on the momentum. The Wigner function depends on both
variables namely on position and momentum. There exists in classical
signal analysis analogous function depending on  time and frequency
introduced  by Ville~\cite{Ville} and it has the properties of quasi-distributions
determined in time-frequency plane. All the discussed functions are
related each to the others by means of different integral transforms The
set of such transforms was studied by Cohen~\cite{Cohen66,Cohen89}. The
transforms are studied e.g., in~\cite{Wolfbook}. One of the integral
transforms is the Radon transform \cite{Radon1917}. It was shown
\cite{TombesiMankoPhys LettA1996} that the Radon transform of the Wigner -Ville function
yields the probability density called quantum tomogram and this tomogram
determines the Wigner-Ville function.

At present, spectacular progress is achieved in the development of probabilistic approach to the basic principles of quantum mechanics (for instance, see \cite{M}). This approach is based on tomographic representation of density matrix and the Wigner function, yielded by the Radon transform. The probabilistic approach made it possible to apply the methods of quantum mechanics in description of various signals. The most comprehensive description of this method is given in \cite{MMM}. In the last decade, the use of tomographic representation is widely employed to reduce noise in severely degraded signals \cite{AM,MMA,AFE}. Especially interesting are the attempts to use the tomographic approach to transform and analyze the medical signals \cite{Aguirre11,Aguirre13}. Processing of such signals widely use the Fourier transform, which is now a routine tool in interpretation of electrocardiograms (ECG) (for example, see \cite{Revenko2015, Mikhailov2017}). The mathematical apparatus of quantum mechanics operates with vectors in Hilbert space or complex wave functions. In contrast, the medical signals are variable physical quantities described by the real functions. Therefore, the first section of this paper close the gap between the concepts of quantum mechanics and the real physical signals such as ECG. The second section explains the tomography mathematics, and the third one applies it to analyze two typical ECG led from a healthy examinee and a patient with the early signs of IHD.
It is reasonable to remember that the early diagnostics of cardiovascular diseases is an urgent medical and social problem, which became increasingly important in relation to unfavorable changes in the modern lifestyle manifested by dramatic decrease of physical load and exercises, imbalanced diet, and augmented stress originated from the modern multimedia environment. It is a common knowledge that numerous severe pathologies (the cardiovascular included) can be coped at the initial stage provided the correct diagnosis was established with the early diagnostic methods. The problems of such diagnostic relate to similarity between the parameters of the patients with an early developing pathology and those of a healthy person. This study was designed to assess the novel vistas opened by analysis of human ECG with its tomographic representation.

\section{Signal analysis}

Normal signal of any nature (e.g. electromagnetic, acoustic, seismic, the signal on the situation in the financial markets and so on) is a physical value, which is set by a real time-dependent function  $f(t)=f^*(t)$ as usual. Different integral transformations are used when a signal is processed. Among the Fourier and wavelet linear transformations like are the more developed. We obtain a signal representation on a frequency axe if the Fourier-transformation of a signal is done. In this case an examined signal will be represented as a complex function of a frequency  $f(\omega)$ in a general situation. In other words, we can represent the same physical value but in different ``frame systems'', ie. in different representations. From this point of view there is a complete analogy with a quantum system state $\psi$ description in different representations. We know that a quantum system state is given by a Dirac vector $|\psi\rangle$ in a Hilbert space, which is determined by a wave function  $\psi(x)=\langle x| \psi\rangle$ in a coordinate space (see, eg review \cite{mendes2015}). In a momentum space this state is given by a function of momentum $\psi(p) \equiv \psi_p=\langle p| \psi\rangle$ respectively.

Let consider signals which are given by vectors in a Hilbert space $|f\rangle$, and a dependence on time of this signal is nothing else but a temporary representation of that, i.e.
\begin{equation}\label{1s}
  f(t)=\langle t|f\rangle.
\end{equation}

The relation  \eqref{1s} is valid when a set of vectors  $|t\rangle$ is the basis. So, if we continue this analogy we need to introduce a {\it Hermitian} operator of time $\hat t$ which is analogous to the space coordinate operator  $\hat q$ in quantum mechanics\footnote{In other words, we consider a signal description in ``$t$-representation''.} It means that we are introducing a formalism of signal vector states $|f\rangle$ in $t$-representation:
\begin{equation}\label{2s}
  {\hat t}f(t)=tf(t),
\end{equation}
where $t$ is the current time.

We can introduce a {\it Hermitian} operator of frequency $\what{\omega}$ because consider both a temporary and a frequency representation of a signal. In correspondence with the Fourier transformation it acts on the signal function in a time representation as follows
\begin{equation}\label{3s}
  {\what \omega}f(t)=-{\i}\frac{{\p}f(t)}{{\p}t}.
\end{equation}
The time and frequency operators introducing by this way satisfies the following commutation relation
\begin{equation}\label{4s}
  [{\hat t},{\what \omega}]={\i}.
\end{equation}
This relation can be considered as an analogue of the commutation relation for operators of coordinates and momentums for quantum systems. So, corresponding to the operators  $\hat t$ and $\what \omega$ variables are canonically conjugate.

The so called {\it ``analytical signal''} $f_a(t)=\langle t|f_a\rangle$ is putted in line to a real signal $f(t)$  in accordance with the following rule. Let us consider a Fourier transformation of a real function\footnote{Everywhere further an integral written without limits understood as a definite integral in infinite limits   $-\infty <t<\infty$.}
\begin{equation}\label{5s}
  \langle\omega|f\rangle=\frac{1}{2\pi}\int{\d}tf(t){\e}^{-{\i}\omega
  t}\equiv \int{\d}t\langle\omega|t\rangle\langle t|f\rangle.
\end{equation}
We can interpret a spectral decomposition as ``a frequency representation'' of a signal $|f\rangle$. We have identified, however {\it a transition matrix}, which is an analogue of the de Broglie wave
\begin{equation}\label{6s}
 \langle t|\omega\rangle=
 \frac{1}{\sqrt{2\pi}}{\e}^{{\i}\omega t}.
\end{equation}

A complex analytical signal $f_a(t)$ is determined by ``a wave function'' in a frequency representation as follows
\begin{equation}\label{7s}
 \langle\omega|f_a\rangle = \langle\omega|f\rangle\theta(\omega).
\end{equation}
Here is the Heaviside function
\begin{equation}\label{8s}
  \theta(\omega)=\frac{1}{2}\left(1+\frac{|\omega|}{\omega}\right)=
  \left\{
  \begin{array}{cl}
    1, \quad & \omega >0, \\
    0, \quad & \omega <0.
  \end{array}
  \right.
\end{equation}

So, Fourier components for $\omega >0$ of a Fourier decomposition of an analytical signal $f_a(t)$ coincide with Fourier components of a real signal $f(t)$. For negative frequencies $\omega <0$ they are equal to zero.

``The wave function'' of an analytical signal $|f_a\rangle$ in a time-representation is determined as follows
\begin{align}\label{9s}
  \langle t|f_a\rangle =f_a(t)=&\,
  \frac{1}{\sqrt{2\pi}}\int{\d}\omega
  \langle\omega|f\rangle\theta(\omega){\e}^{{\i}\omega
  t}=\nonumber \\
 =&\, \frac{1}{\sqrt{2\pi}}\int\limits_0^{\infty}{\d}\omega
  \langle\omega|f\rangle{\e}^{{\i}\omega t}.
\end{align}
In other words we can write
\begin{equation}\label{10s}
  f_a(t)=\int\limits_{-\infty}^{\infty}{\d}t'
  \int\limits_0^{\infty}\frac{{\d}\omega}{2\pi}
  f(t'){\e}^{{\i}\omega (t-t')}.
\end{equation}
Let us use the well known relation for generalized functions
\begin{equation}\label{11s}
  \delta_-(t)=\int\limits_0^{\infty}\frac{{\d}\omega}{2\pi}
  {\e}^{{\i}\omega t}=\frac{1}{2}\delta(t)+
  \frac{\i}{\pi}\wp \frac{1}{t},
\end{equation}
where the symbol $\wp$ indicates the main value. So, the application of $\wp t^{-1}$ to the integral of any function $\Phi(t)$ understands as follows
\begin{equation*}
  \wp \int\limits_{-\infty}^{\infty}{\d}t\Phi(t)\frac{1}{t}=
  \lim_{\tau \to 0}\left[\int\limits_{-\infty}^{-\tau}
  \Phi(t)\frac{{\d}t}{t}
 +\int\limits_{\tau}^{\infty}\Phi(t)\frac{{\d}t}{t}\right].
\end{equation*}
Application of the property \eqref{11s} to a definition of the analytical signal \eqref{10s} shows that $f_a(t)$ has an imaginary part. So, it is connected with a real signal by the following relation
\begin{equation}\label{12s}
  f_a(t)=\frac{1}{2}f(t)-\frac{\i}{\pi}
  \wp \int\limits_{-\infty}^{\infty}{\d}t'\frac{f(t')}{t-t'}.
\end{equation}
We can find an analytical signal $f_a(t)$ if a real signal $f(t)$ is known and backwards:
\begin{equation}\label{13s}
  f(t)=f_a(t)+f_a^*(t),
\end{equation}
where $f_a^*(t)$ is a complex conjugate function.

Note that both real and imaginary part of an analytical signal are connected with each other by a ``dispersion relation'' which is identical to the well known dispersion relation for susceptibilities in statistical physics and macroscopic electrodynamics.

We can change the order of integration in the analytical signal definition \eqref{10s}. Really, let us use the definition \eqref{10s} in Eq.\eqref{13s} taking into account that a signal function $f(t)$ is real:
\begin{align*}
  f_a(t)=&\int\limits_{-\infty}^{\infty}{\d}t'
  \int\limits_0^{\infty}\frac{{\d}\omega}{2\pi}\bigl(
  f_a(t')+f_a^*(t')\bigr){\e}^{{\i}\omega (t-t')} =\\
  =&\int\limits_{-\infty}^{\infty}{\d}t'
 \int\limits_{-\infty}^{\infty}\frac{{\d}\omega}{2\pi}
  f_a(t'){\e}^{{\i}\omega (t-t')}.
\end{align*}
Thus we have
\begin{equation}\label{14s}
 \frac{1}{2\pi}\int\limits_0^{\infty}{\d}\omega
 \left[\int\limits_{-\infty}^{\infty}{\d}t'
 f_a(t'){\e}^{{\i}\omega (t-t')}\right]=f_a(t),
\end{equation}
Eq.\eqref{14s} shows that both integrals of an analytical signal with generalized function $\delta_-(t-t')$ and with ``complete''  $\delta$-function coincide.

\section{Density matrix, the Wigner function and a tomogram of a signal}

The approach to signal analysis given in the previous section allow us to continue analogies with quantum systems description. Namely, we will apply a tomography representation for signal description. Thus, one can assume that totality of all signal values is represented by a set of vectors $|f\rangle$ in the space of all possible values. Symbol $f$ designates a signal amplitude in this case. This value can be a complex one in general. The space of states must be complete, ie.
\begin{equation}\label{s15}
  \langle f'|f\rangle = \delta(f-f') \quad
  \mbox{and} \quad  \int {\d} f |f\rangle\langle f| = \hat{1}.
\end{equation}
This assumes that a signal amplitude can take continuous values in some interval  $[f_{\rm min}, f_{\rm max}]$. In other words, it has a continuous spectrum. A signal takes a discrete values if it is represented in a digital form. In this case properties \eqref{s15} must be rewritten as follows
\begin{equation}\label{s16}
  \langle f'|f\rangle = \delta_{f,f'} \quad
  \mbox{and} \quad  \sum_{f=f_{\rm min}}^{f_{\rm max}} |f\rangle\langle f| = \hat{1}.
\end{equation}
Let us introduce a density matrix of a signal using quantum mechanics methods. It will be determined as a density matrix of a pure state:
\begin{equation}\label{s17}
  \rho^s= |f\rangle\langle f|.
\end{equation}
This relation in a ``time-representation'' has the following expression:
\begin{equation}\label{s18}
  \rho^s (t,t')= \langle t|f\rangle\langle f|t'\rangle \equiv
   f(t)f^{*}(t').
\end{equation}
Let define the Wigner-Ville function of a signal as follows \cite{Ville}
\begin{equation}\label{s19}
  W^s(\omega,t)= \int\limits_{-\infty}^{+\infty}{\d}\, x
  \rho^s\left(t+\frac{x}{2}, t-\frac{x}{2}\right)
  {\e}^{-{\i}\omega x}.
\end{equation}
The reverse transformation is determined in  complete correspondence with the Wigner function of a quantum system:
\begin{equation}\label{s20}
  \rho^s(t,t')=
  \int\limits_{-\infty}^{+\infty}\frac{{\d}\omega}{2\pi}
  W^s\left(\omega,\frac{t+t'}{2}\right){\e}^{{\i}\omega(t-t')}.
\end{equation}
A signal can be expressed through the Wigner function using Eq.\eqref{s18} for $t'=0$:
\begin{equation}\label{s21}
  f(t)f^{*}(0)=
  \int\limits_{-\infty}^{+\infty}\frac{{\d}\,\omega}{2\pi}
  W^s\left(\omega,\frac{t}{2}\right){\e}^{{\i}\omega t}.
\end{equation}
The value of a signal at the initial time in any case can be represented as follows
\[
  f^*(0)= |f(0)|{\e}^{-{\i}\varphi(0)},
\]
where a phase $\varphi(0)$ at the initial time remains undetermined. Accordingly with it we have
\begin{equation}\label{s22}
  f(t)= {\e}^{{\i}\varphi(0)}
  \frac{\frac{1}{2\pi}\int\limits_{-\infty}^{+\infty}{\d}\omega
  W^s\left(\omega,\frac{t}{2}\right){\e}^{{\i}\omega t}}
  {\sqrt{\frac{1}{2\pi}\int\limits_{-\infty}^{+\infty}{\d}\omega
  W^s(\omega,0)}}.
\end{equation}

Now we can define a signal tomogram \cite{MM}
\begin{equation}\label{s23}
  w^s(X,\mu,\nu)=\iiint\limits_{-\infty}^{\infty}
  \frac{{\d}t{\d}\omega{\d}u}{(2\pi)^2}
  W^s(\omega,t){\e}^{-{\i}u(X-\mu t -\nu \omega)}
\end{equation}
Here $X$ is an eigen value of a Hermitian operator
\[
  \widehat{X}= \mu \hat{t} +\nu \widehat{\omega}, \quad \mu^2 +\nu^2 =1.
\]
One takes as usual
\begin{equation}\label{s24}
    \mu=\cos\theta, \, \nu=\sin\theta, \quad 0\leq \theta \leq \pi/2.
\end{equation}

The eigen value of $\widehat X$ is located at the straight line of the first quadrant in the plane of variables $(t,\omega)$.

The Wigner unction can be written down explicitly through a signal function after substituting Eq.\eqref{s23} into the definition \eqref{s19}
\begin{equation}\label{s25}
  w^s(X,\mu,\nu)=\iiiint \frac{{\d}t{\d}\omega{\d}u {\d}y}{(2\pi)^2}
  f(t+y/2)f^{*}(t-y/2){\e}^{-{\i}\omega y}{\e}^{-{\i}u(X-\mu t -\nu \omega)}.
\end{equation}
One can integrate Eq.\eqref{s25}  over the $\omega$ variable
\begin{equation}\label{s26}
  \int\limits_{-\infty}^{\infty} {\d}\omega {\e}^{-{\i}\omega(y-\nu u)}=
  2\pi \delta(y-\nu u).
\end{equation}
Thus we have
\begin{equation}\label{s27}
  w^s(X,\mu,\nu)=\frac{1}{2\pi}\iiint {\d}t{\d}u {\d}y
  f(t+y/2)f^{*}(t-y/2){\e}^{-{\i}u(X-\mu t)}\delta(y-\nu u).
\end{equation}
Performing the integration over $u$ we obtain
\begin{equation}\label{s28}
  w^s(X,\mu,\nu)=\frac{1}{2\pi|\nu|}\iint {\d}t {\d}y
  f(t+y/2)f^{*}(t-y/2){\e}^{-{\i}\frac{y}{\nu}(X-\mu t)}.
\end{equation}
Let us change variables in the integral \eqref{s28} as follows
\begin{equation}\label{s29}
  t=(z+q)/2, \quad y=z-q.
\end{equation}
The Jacobian of the transformation \eqref{s29} is equal to -1, but the integration limits extend from
$-\infty$ to $+\infty$ and we obtain
\begin{align}\label{s30}
  w^s(X,\mu,\nu)=&\frac{1}{2\pi|\nu|}\iint {\d}z {\d}q
  f(z)f^{*}(q){\e}^{-{\i}\frac{1}{\nu}
  \left((z-q)X-\mu (z^2-q^2)/2\right)}= \nonumber\\
  = &\frac{1}{2\pi|\nu|}\left|\int\limits_{-\infty}^{\infty} {\d}z f(z)
  {\e}^{\frac{{\i}\mu}{2\nu}z^2 - \frac{{\i}X z}{\nu}}\right|^2.
\end{align}

Finally, let us rewrite a tomogram definition introducing more ``usual'' designation of the integration variable by letter $t$ as follows
\begin{equation}\label{s31}
  w^s(X,\mu,\nu)=\frac{1}{2\pi|\nu|}
  \left|\int\limits_{-\infty}^{\infty}
  f(t)\exp\left[\frac{{\i}\mu}{2\nu}t^2 -
  \frac{{\i}X t}{\nu}\right]{\d}t \right|^2.
\end{equation}

The obtained tomographic representation is strictly connected with the so called fractional Fourier transformation (FRFT) (see, eg. \cite{namias}), which is widely used for example in quantum optics. From the other side it is the same as the Radon transformation (\cite{AdImElPh}). As usual FRFT is written in a form
\begin{equation}
\label{s32}
F(X,\alpha) = \sqrt{\frac{\nu-i\mu}{2\pi\nu}} \exp \left( \frac{i \mu}{2 \nu} X^2 \right) \int\limits_{-\infty}^{\infty} \exp \left[ \frac{i}{2\nu}
\left( -2Xt+\mu t^2 \right) \right] f(t) dt.
\end{equation}
Dimensionless variable $\alpha$ changes in the interval $0 \le \alpha \le 1$ and is called as the transformation order. The transformation order is connected with the previously introduced parameters by the relation $\alpha=2\theta/\pi$. In this case relation \eqref{s24} can be rewritten a follows
\begin{equation}\label{32s}
\mu=\cos\frac{\pi}{2}\alpha, \quad
\nu=\sin\frac{\pi}{2}\alpha
\end{equation}

Easy to see that the tomographic representation is equal to the squared of FRFT modulus
\begin{equation}\label{s33}
    w^s(X,\mu,\nu)=|F(X,\alpha)|^2
\end{equation}

\section{ Tomogram calculation}

The identical transformation is realized at $\alpha=0$ and at $\alpha=1$ we obtain the Fourier transformation. Numerical calculation on a computer of this kind of transformation is complicated by two reasons. At first, the real signal (cardiogram) has a complex fine-scale structure and demands large number of a grid nodes in a numerical representation. At second, the integral transformation core is an imaginary exponent of a quadratic function. So, it is a frequently oscillating function and gives additional problems in approximation of the integral by quadrature formulae. The best way to overcome these difficulties is in a reduction of FRFT to the integral of the convolution type (see, eg. \cite{ozaktas}). The following application of the fast Fourier transformation algorithm makes the procedure more successful. Using the identity  $\mu X^2 -2Xt +\mu t^2 = (\mu-1) X^2 +(X-t)^2 +(\mu-1) t^2$ \cite{ozaktas} one can transform Eq.\eqref{s32} in the following form

\begin{equation}
\label{s34}
F(X,\alpha) = \sqrt{\frac{\nu-i\mu}{2\pi\nu}} \exp \left( \frac{i (\mu-1)}{2 \nu} X^2 \right) \int\limits_{-\infty}^{\infty} \exp \left[ \frac{i}{2\nu}
\left( X-t \right)^2 \right] \tilde f(t) dt.
\end{equation}
Here is
\[
 \tilde f(t) =  \left( \frac{i (\mu-1)}{2 \nu} t^2 \right) f(t).
\]
Integral in Eq.\eqref{s34} has a type of convolution.

Interval $0 \le \alpha \le 1$ was broken for 50 equal distances and parameter $\alpha$ taken 51 values including the boundaries of the interval. Quadrature formula for integral \eqref{s34} was constructed on a uniform grid. The grid size was determined automatically for the given precision of calculations. Permissible relative error of calculations was taken to be of the order of $10^{-5}$. This precision was achieved with the grid size up to $2^{24}$ knots. One tomogram calculation on PC was executed from few to tens minutes of a computer time due to the FFT algorithm application.

It should be noted that a tomogram form essentially depends on a choice of a unit of time representation. In primary data received from a device time $T$ is given in seconds as usual. When a tomogram is calculated a dimensionless time  $t=T/T_0$ is used, where $T_0$ is a unit of time in seconds. A frequency scale in the tomogram cross-section at $\alpha=1$ is determined unambiguously by a choice of $T_0$. Namely, a value $\omega=T_0^{-1}X$ is a circular frequency and it has a dimension of rad/sec.

\section{Tomograms for ECG-signal}

The obtained results were applied for processing of human electro-cardiogram signals (ECG). For these reasons two ECG with high resolution was taken from the data base of medical center ``Practical Neurology'' , Moscow. One of them was recorded from a healthy man aged 27 years, and the second from a patient (aged 70 years) with the early ischemic heart disease (IHD) manifested by a mild chest pain during physical exercises.

ECG were led in supine position under normal breathing with original electrocardiography amplifier characterized by the input short-circuit noise of 3600 and 900 nV2/Hz in the frequency bands of 0.1-0.2 and 1.0-2.0 Hz, correspondingly. Resolution was further enhanced by a 100-fold decimation of primary digital signal, which reduced the sampling rate to 160 Hz and decreased the input noise of instrumentation system by 10 times \cite{Revenko2015}.

A differential ECG-signal was led with two chest electrodes $V_1$ and $V_2$ made of Ag/AgCl. The resulting signal was equal to the difference $(V_1-V_2)$. The reference electrode was mounted on the right leg. This recording mode with a short interelectrode distance (about 10 cm) diminished oscillations of the isoline resulted from the mechanical effect of respiration excursions.

\begin{figure}
\begin{minipage}{0.48\linewidth}
\includegraphics[width={1.0\linewidth}]{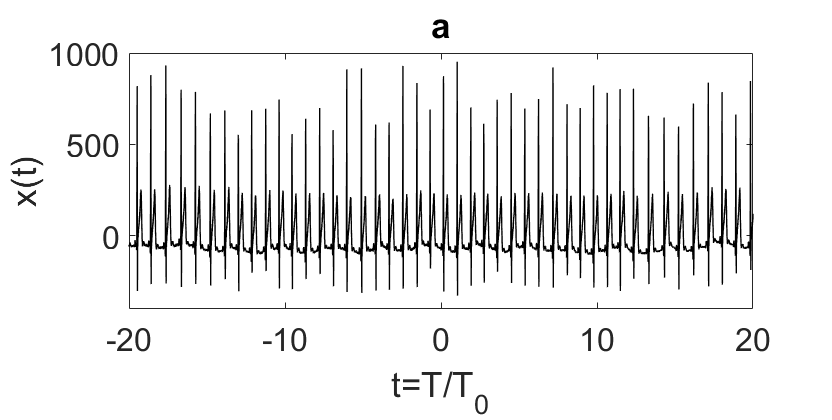}
\end{minipage}
\hfill
\begin{minipage}{0.48\linewidth}
\includegraphics[width={1.0\linewidth}]{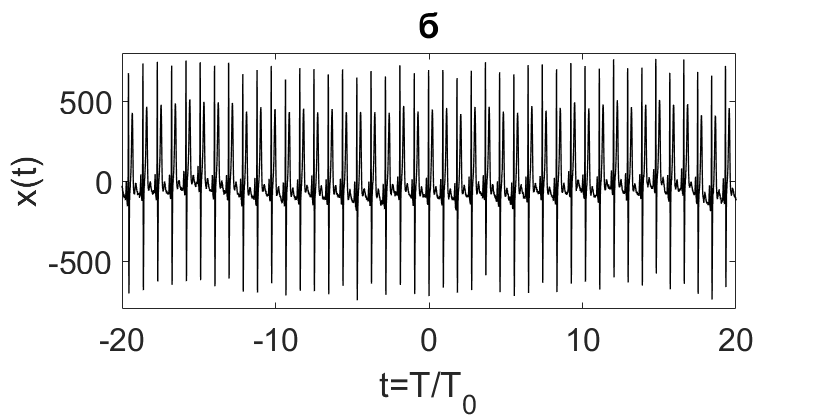}
\end{minipage}
\caption{Analyzed ECG segments of a healthy man {\it a}) and a patient with initial symptoms of IHD {\it b}). Ordinate in microvolt. Time unit is $T_0=1$ sec.}  \label{fig:long}
\end{figure}

Fig.\ref{fig:long} shows both analyzed ECG. It clearly demonstrates unobviousness of direct ``visual'' assessment of the heart state by ECG shape at initial stages of cardiac pathology. Really, the evident ``instability'' of ECG amplitude (Fig.\ref{fig:long}{\it a}) reflects its normal variations during the respiratory cycle, including those related to neural cardiotropic influences. In contrast, the more stable form of ECG in IHD patient  Fig.\ref{fig:long}{\it b}  can result from pathological alteration in the heart manifested by abnormal (enhanced) automaticity \cite{Janse MJ}.

We limited the analyses of ECG by the segment length of 40 sec, which included about 40 cardiac cycles. This length encompassed the slow oscillations of ECG at the frequencies related to Mayer waves (about 0.1 Hz) \cite{Julien2006} and respiration (0.2-0.5 Hz).


\begin{figure}
 \centering
 \includegraphics[height=5cm]{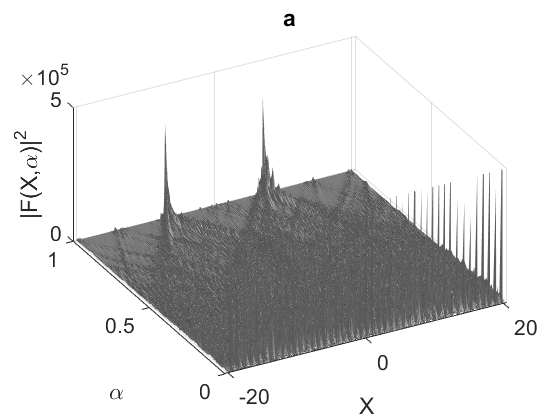}
 \caption{A tomogram of ECG signals shown at Fig.\ref{fig:long} for a healthy men {\it  a})}\label{fig:long-Rh}
\end{figure}

\begin{figure}
 \centering
 \includegraphics[height=5cm]{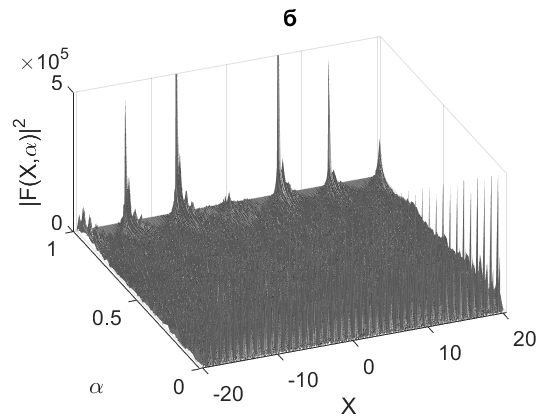}
 \caption{A tomogram of ECG signals shown at Fig.\ref{fig:long} for a patient with initial symptoms of IHD {\it b})}\label{fig:long-Rp}
\end{figure}

Let we note, at first, that a tomogram can take nonnegative values only, and because of this the square of a signal modulus will be given at the following figures. A tomogram of a signal allows us to track a continuous transition from a temporary representation to its Fourier spectrum.  Since tomogram is a function of two variables it can be represented by a 3D plot. This graphic contains a large information and it is very complicated for analysis in a general case. So, examination of a cross-section at different values of a transformation order ( $\alpha$-angle) is more suitable. Nevertheless, the complete tomogram of the signals are shown at Figs.\ref{fig:long-Rh} and \ref{fig:long-Rp} for the illustration of the total picture. A tomogram shows that a Fourier spectrum is symmetrical because the frequency varies in infinite limits  $-\infty <\omega <\infty$ in representation  \eqref{s23}.

Visual analysis of three-dimensional pictures shows very relief pattern with a large number of peaks and the number of peaks in a healthy patient is less with respect to the patient with IHD. One can see that maximum of a mixture of two  signal representation at the ``angle'' $\alpha = 1/\sqrt{2}$ is achieved and a tomogram is the least informative. The picture essentially changes when approaching  $\alpha \to 1$. The Fourier spectrum formation strongly depends on details of the initial time-representation of a signal in this case. Note also, that an essential influence on processing results has an existence of a linear changing of electrode potentials at the time interval under examination because of parasitic contribution to the Fourier spectrum. It is necessary to correct a linear ``trend'' if it exists before the beginning of a tomogram calculation.

\begin{figure}
\begin{minipage}{0.48\linewidth}
\includegraphics[width={1.0\linewidth}]{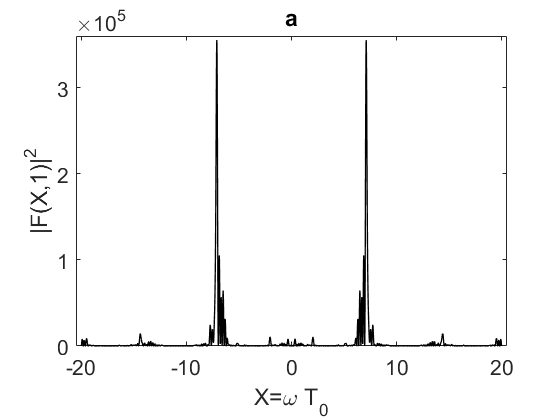}
\end{minipage}
\hfill
\begin{minipage}{0.48\linewidth}
\includegraphics[width={1.0\linewidth}]{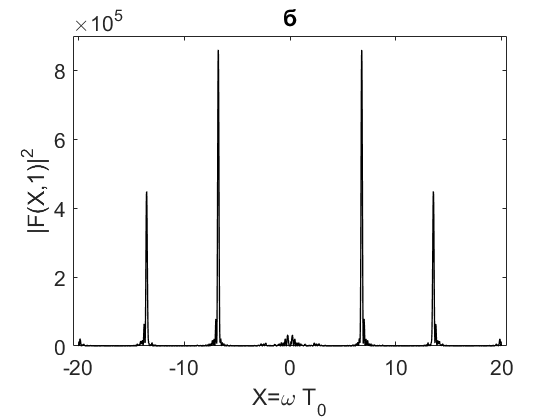}
\end{minipage}
\caption{Square of the Fourier transformation modulus of ECG-signal, shown at Fig.\ref{fig:long} for a healthy men {\it  a}) and for a man with initial symptoms of IHD {\it b})} \label{fig:long-f}
\end{figure}

\begin{figure}
\begin{minipage}{0.48\linewidth}
\includegraphics[width={1.0\linewidth}]{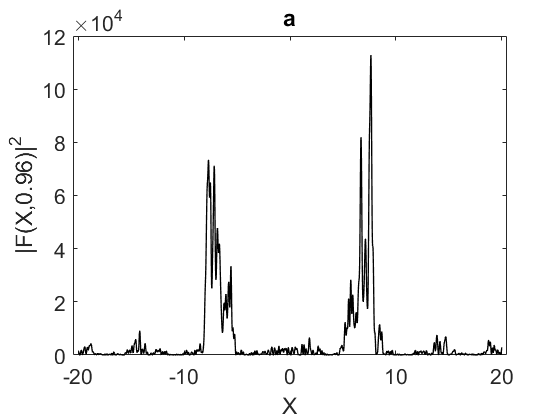}
\end{minipage}
\hfill
\begin{minipage}{0.48\linewidth}
\includegraphics[width={1.0\linewidth}]{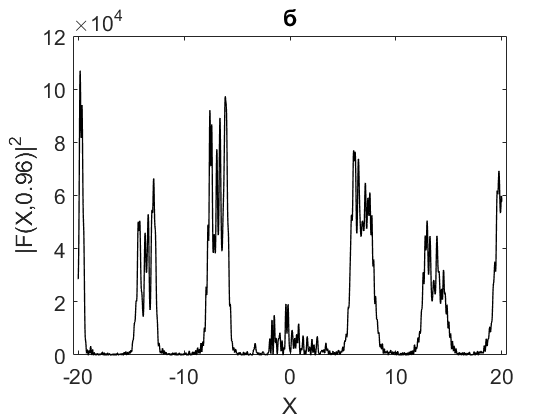}
\end{minipage}
\caption{Cross-section of ECG-signal tomogram shown at Fig.\ref{fig:long} for transformation parameter  $\alpha=0.96$. {\it a}) -- for a healthy men and {\it b}) for a man with initial symptoms of IHD.} \label{fig:long-s}
\end{figure}

Visual inspection of 3D tomograms (Fig.\ref{fig:long-Rh} and \ref{fig:long-Rp}) can select the value of $\alpha$, which corresponds to the most prominent differences between two 3D plots manifested by their peaks. Evidently, two plots are especially different at $\alpha$ approximating to 1. To show these differences most illustratively, the following data were plotted as the Fourier spectrum of the signals under examination at Fig.\ref{fig:long-f} and the cross-section for $\alpha=0.96$ at  Fig.\ref{fig:long-s}. Both figures show that IHD patient had extra well-resolved peaks in comparison with that of healthy person.

This is an expected result because it reflects abnormal (enhanced) automaticity of the heartbeats, which is characteristic of various heart diseases, IHD included \cite{Antoni1992}.

Comparison of the Fourier spectrum at Fig.\ref{fig:long-f} and a cross-section of the tomogram at Fig.\ref{fig:long-s} shows that the latter reveals more differences between both examined ECG signals. Actually, while ECG spectrum of normal examinee did not evidently change after replacing the routine Fourier transform by the cross-section of the tomogram at $\alpha=0.96$ except for widening of the spectral peaks, the spectrum of IHD patient acquired the zero-frequency peak and the third harmonics.

When expecting the extra peaks in Fig.\ref{fig:long-s}, one can see that 3D plot at Fig.\ref{fig:long-Rp} also has the corresponding ``hills'', which are somewhat shifted from the left back plane. Thus, a relatively small deviation from the classical Fourier transform to the cross-section of the tomogram results in pronounced differences in ECG spectra obtained for the normal examinee and a patient with early IHD.

The data obtained suggest that in contrast to the routine Fourier spectrum of ECG (and more so in contrast to time-domain ECG), to the tomogram with $\alpha \lesssim 1$ makes it possible to detect the early differences between the normal and quasi-normal heart actions.

\section{Conclusion}

Present observations show prospectiveness of ECG analysis with tomogram representation or the Radon transform in the representative groups of healthy persons and patients with the early signs of IHD. In contrast to routine Fourier analysis, the Radon transform revealed the extra spectra peaks in IHD patient. Hopefully, this promising result may lead to the development of an effective method to reveal the early cardiac abnormalities based on the Radon transform.

\section*{Acknowledgments}
This research was partially supported by the Government of the Russian Federation (Agreement 05.Y09.21.0018) and the Ministry of Education and Science of the Russian Federation 16.7162.2017/8.9.

\pagebreak

\end{document}